\documentclass[final,3p,times,twocolumn]{elsarticle}

\usepackage{graphicx}
\usepackage{amssymb}
\usepackage{bm}
\usepackage{amsmath}
\usepackage{showkeys}

\journal{Solid State Communications}

\begin{document}

\begin{frontmatter}

\title{Density functional theory studies of interactions of graphene with its environment: substrate, gate dielectric and edge effects}

\author[mer]{Priyamvada Jadaun\corref{footnote1}}
\ead{priyamvada@utexas.edu}
\ead[https://webspace.utexas.edu/~pj3292/]{Web page}

\author[mer]{Bhagawan R. Sahu}
\author[mer]{Leonard F. Register}
\author[mer]{Sanjay K. Banerjee}

\address[mer]{Microelectronics Research Center, The University of Texas, Austin, TX 78758, USA}
\cortext[footnote1]{Corresponding author}

\begin{abstract}
This paper reviews the theoretical work undertaken using density functional theory (DFT) to explore graphene's interactions with its surroundings. We look at the impact of substrates, gate dielectrics and edge effects on the properties of graphene. In particular, we focus on graphene-on-quartz and graphene-on-alumina systems, exploring their energy spectrum and charge distribution. Silicon-terminated quartz is found to not perturb the linear graphene spectrum. On the other hand, oxygen-terminated quartz and both terminations of alumina bond with graphene, leading to the opening of a band gap. Significant charge transfer is seen between the graphene layer and the oxide in the latter cases. Additionally, we review the work of others regarding the effect of various substrates on the electronic properties of graphene. Confining graphene to form nanoribbons also results in the opening of a band gap. The value of the gap is dependent on the edge properties as well as width of the nanoribbon. 
\end{abstract}


\end{frontmatter}

\section{Introduction}
\label{intro}
Graphene is a monolayer of carbon atoms that is arranged in a honeycomb structure with the atoms bonded together by $\text{sp}^2$ hybridized bonds. Graphene is of incredible interest to the solid state community since it displays very high electron mobility, exceeding $15,000 \text{cm}^2/(\text{V-s})$ and tunable gap at the Fermi level along with high crystal quality. The carriers in graphene are governed by relativistic Dirac equation, thereby displaying zero effective mass \cite{novoselov, geim} and are continuously tunable between electrons and holes. There is tremendous promise for building graphene-based, high quality and fast post-silicon devices. A graphene sheet has two crystallographically equivalent atoms in its primitive unit cell. Two bands originating from $\text{p}_\text{z}$ orbitals cross each other exactly at the Fermi energy, at the K and K' points  of the hexagonal reciprocal unit cell, making graphene a zero-gap semiconductor.
\\Despite graphene's intriguing properties, one of the biggest problems with using it for electronic devices is the lack of an energy gap. There exist two accessible ways to lift the degeneracy at graphene's Dirac point. One is to mix or hybridize the electronic states at K and K' by breaking translational symmetry, a feat that can be achieved by confining graphene in space. The second way is to break the sublattice symmetry within the graphene crystalline structure \cite{zhou}. A simple way to render the A and B sublattices inequivalent is to place graphene on a substrate. Additionally, substrates are also an essential component of the proposed graphene based electronic devices. For this reason, it is quite essential to study the interaction of graphene with its substrate and the latter's impact on graphene's properties. Moreover substrates are also used in the manufacture of graphene. A widely used method for graphene growth is the epitaxial growth of graphene on substrates like silicon carbide (SiC), silicon-dioxide ($\text{SiO}_2$) etc. In addition to its electronic structure, substrates are known to have an effect on the spectroscopic properties of graphene and contribute to its surface roughness as well \cite{shemella}.
\\The gate dielectric is an equally important part of a transistor. Apart from its impact on transconductance, subthreshold swing and frequency response, its interaction with graphene also alters the latter's nature and properties. In particular, the carrier mobility of graphene based transistors can be critically altered by the nature of the dielectric layer and graphene-dielectric interactions \cite{liao}. Most models of graphene transistors employ the substrate as a back gate, covered by a dielectric oxide on which graphene is then placed. The gate dielectric could possibly screen graphene from scattering caused by charge impurities in the substrate, giving mobility a boost. Another kind of device architecture uses an independent top gate in addition to the back gate, which necessitates the deposition of a dielectric on top of graphene. In such a scenario, graphene's performance is again severely deteriorated due to Coulomb interactions of graphene's carriers with charge impurities in the deposited dielectric \cite{puls}. It is therefore, quite important to study and understand the impact of dielectrics on graphene's performance.
\\While discussing the properties of an infinite 2 dimensional graphene sheet is essential and quite useful to make practical device applications a reality, actual devices will contain finite sheets of graphene. It thus becomes necessary to look into the behavior of confined graphene sheets such as graphene nanoribbons (GNRs). Besides, confinement of graphene is a good way of introducing a band gap in the material which is, in turn, critical for transistor operation. With confinement, edge effects also start playing a role and one needs to study how they influence the properties of graphene as well. In this paper, we review density functional theory (DFT) studies of graphene's interactions with its environment, i.e., substrates and dielectrics, as well the impact of edges on graphene's properties. In particular, we focus on our own work \cite{jadaun} on graphene's interactions with two oxides, i.e.  alpha-quartz ($\text{SiO}_2$) and alpha-sapphire or alumina ($\text{Al}_2\text{O}_3$).

\section{Graphene's interaction with substrates}
\label{substrate}

\begin{figure}[ht!]
\scalebox{0.18}{\includegraphics{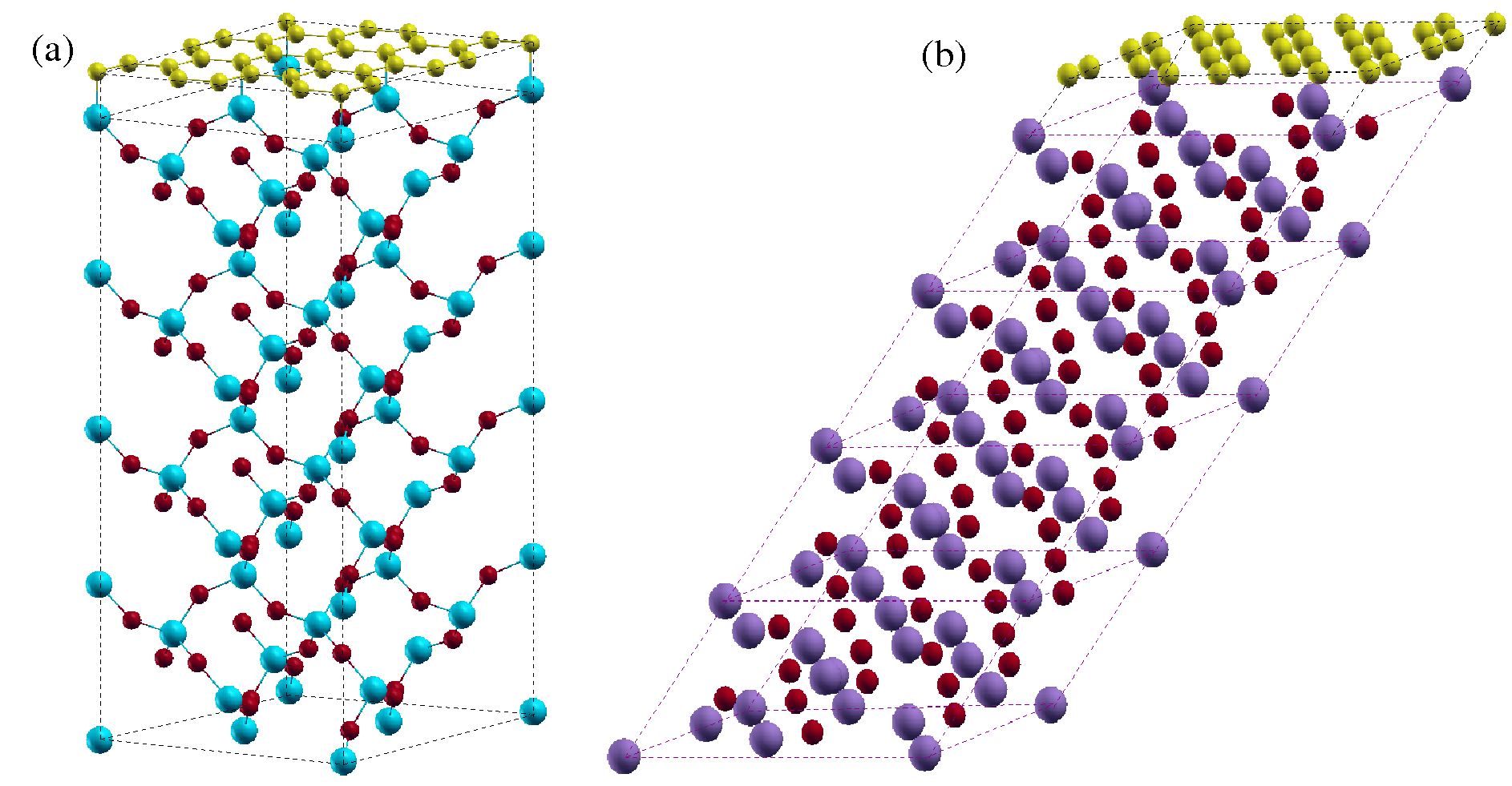}}
\caption{(Color online) Schematic illustrations of the supercell structure of monolayer graphene on (a) Si-terminated quartz (b) Al-terminated alumina. The atoms are shown in color: Si (blue), Al (purple), O (red) and C (yellow).\label{fig1}} 
\end{figure}

Silicon carbide (SiC) is one of the most popular substrates used for graphene fabrication. Varchon et. al. \cite{varchon} performed DFT calculations of graphene on (0001) and ($000\bar{1}$) 4H-SiC (silicon and carbon terminated, respectively) surfaces. Graphene is almost commensurate with these SiC surfaces with a common cell corresponding to a $6\sqrt{3}*6\sqrt{3}R30$ reconstruction (with respect to a 1x1 SiC surface unit cell). The actual structures used in this calculation were the smaller $\sqrt{3}*\sqrt{3}R30$ cell which corresponds to a 2x2 graphene unit cell. The latter approximation necessitated an 8\% stretch in the graphene sheet for it to match the SiC lattice parameter. After relaxation, they found that the first graphene layer, immediately next to the SiC surface, lay at a distance of 2.0$\AA$ for the Si-terminated surface and 1.66$\AA$ for the C-terminated surface and displayed strong chemical interactions with both. This covalent bonding completely destroyed the linear band dispersion of that first layer, opening a large band gap. The Fermi level was pinned by a state with small dispersion in real space, which arose from the dangling bonds at the SiC terminating surface. Interestingly they also found that the final positions of the bulk layers depended only on the first carbon layer. An addition of graphene layers above the first one did not alter the structure of SiC or the first graphene layer. The second graphene layer lay 3.8$\AA$ approx. above the first one with subsequent planes spaced by 3.9$\AA$. Except for the first layer, the higher graphitic layers were bound by weak van der Waals forces only. Graphene related dispersions were recovered upon adding more carbon layer(s) to the structure. This showed that the first layer shielded further layers from interactions with the underlying substrate, thereby earning it the name of the buffer layer. The Si-terminated surface had the Fermi level falling 0.4 eV above the Dirac point, doping the graphene sheets \textit{n}-type. However C-terminated surfaces showed neutral graphene sheets with Fermi level lying on the Dirac point. Clear signs of charge transfer between the substrate and the graphene buffer layer were also seen, displaying typical characteristics of convalent bonding. Higher graphene sheets showed more delocalised charge density profiles.
\\Another DFT study of graphene (monolayer and bilayer) on 6H SiC surface \cite{matta} used a unit cell of graphene on a $\sqrt{3}*\sqrt{3}R30$ reconstruction of the SiC surface unit cell. The supercell used for calculation used 6 bilayers of SiC with single or bilayer graphene. While the reconstruction again caused an 8\% mismatch of lattice constants between SiC and graphene, after relaxation the elastic energy was significantly reduced. The unit cell at the interface consisted of three substrate surface atoms and four graphene unit cells. The dangling bonds at the corners of the unit cell remained unsaturated while the others bonded to graphene atoms. The equilibrium distance of graphene from Si-terminated surface was 2.58$\AA$ while its distance from C-terminated surface was 2.44$\AA$. Both the terminations were binding, releasing 0.72eV (0.60 eV) of energy per graphene unit cell for the Si-terminated (C-terminated) surface. For both surfaces the substrate bonding atoms relaxed towards graphene while the corresponding graphene atom relaxed towards the substrate surface, reducing their distance to enable them to bond. The graphene layer was more rigid than the underlying substrate and actually distorted the substrate lattice, contrary to conventional adsorption models. This hinted at the strong in-plane bonding within the graphene layer which overshadowed the coupling to the substrate surface. When a second graphene layer was added, it relaxed to a separation of 3.3$\AA$, which is quite close to the bulk graphite value of 3.35$\AA$. The first layer of graphene was seen to act as a buffer for the second layer, much like the case before, protecting it from interactions with the substrate. The covalent bonding between the buffer graphene layer and the substrate again drastically altered monolayer graphene's band spectrum. The Dirac cones combined with the SiC valence band and upper graphene bands merged with the conduction band leaving a band gap in between. There existed a localised interface state arising from graphene interactions with the substrate dangling bonds. While two of the three dangling bonds interacted with graphene, the third one remained unsaturated. The Si-terminated structure was found to be half-filled metallic while the C-termination was insulating. A study of charge density profiles revealed that on Si-terminated SiC, the interface state was strongly delocalised due to hybridisation with graphene states in the conduction band. The C-terminated surface however retained its localised character leading to spin polarisation and thus opening of a gap. On Si-terminated SiC the second graphene layer tended to be \textit{n}-doped while it was neutral on C-terminated SiC. 
\\Xu et. al. \cite{xu} conducted a study on graphene's interaction with Si(100) substrate. They performed DFT calculations on graphene flakes and graphene sheets with hydrogen (H) passivated edges placed on Si(100) surfaces. In the case where Si dangling bonds were passivated with H, there was no bonding between graphene flake and the Si surface. This meant that graphene's Dirac spectrum remained undisturbed. This was evident from the large equilibrium distance between graphene and the Si surface as well as the lack of charge sharing between Si and graphene. However once H was desorbed from the Si surface, the surface Si atoms relaxed outwards and graphene relaxed towards the Si surface. Covalent bonding between C and Si occured leading to Si-C bond pairs that ranged in length from 1.93$\AA$ to 2.05$\AA$. These were found to be quite close to the SiC bond length of 1.92$\AA$, pointing to the strong nature of covalent interaction between graphene and the substrate. Due to strong localisation of electron density in the latter case, the gap seen in graphene's sectrum was also much larger than the previous graphene on Si(100)-H termination. Similar Si-C bond formation in the Si(100) unpassivated termination was seen for infinite graphene sheets. The surface state electrons were seen to transfer from Si to the graphene sheet leaving it \textit{n} doped. 
\\Hexagonal boron nitride (\textit{h}-BN) can be treated as either substrate or dielectric for graphene based devices. Hexagonal BN is a wide gap crystalline insulator (gap of 5.97 eV) with a very small lattice mismatch with graphene i.e. only 2\%. It has a layered structure which resembles that of graphene except that in \textit{h}-BN the two sublattices are inequivalent. As in graphite, the interaction between adjacent layers in BN is weak and the stacking is AB type. Local density approximation (LDA) gives the layer separation to be 3.24$\AA$ \cite{giovannetti} which is quite close to the experimentally observed value of 3.33$\AA$. In this calculation a composite lattice constant of 2.445$\AA$ was used for the entire \textit{h}-BN and graphene structure. Three distinct types of interfaces were considered namely, (a) one C atom over boron(B) and the other over nitrogen(N); (b) one C atom over N and another above the center of BN hexagon and (c) one C atom over B while the other above the center of the BN hexagon. Self-consistent calculations revealed the lowest energy and hence the most stable structure to be (c) with one C above B and another above the BN hexagonal center. The equilibrium separation between graphene and \textit{h}-BN for case (c) was 3.22$\AA$ followed by 3.40$\AA$ for case (b) and finally 3.50$\AA$ for case (a). Electronic band structures were subsequently computed for the lowest energy structures in all cases. Hexagonal-BN displayed a band gap of 4.7 eV which is quite close to the bulk value. This hinted at the weak nature of the interaction of \textit{h}-BN with graphene. Graphene's Dirac cone displayed the opening of a gap, as expected. A gap of 53 meV was seen for the stablest configuration (c). There were similar gap openings seen for (a) and (b), i.e. 56 meV and 46 meV, respectively. 

\section{Impact of dielectrics on graphene}
\label{dielec}

\begin{figure}[ht!]
\scalebox{0.3}{\includegraphics{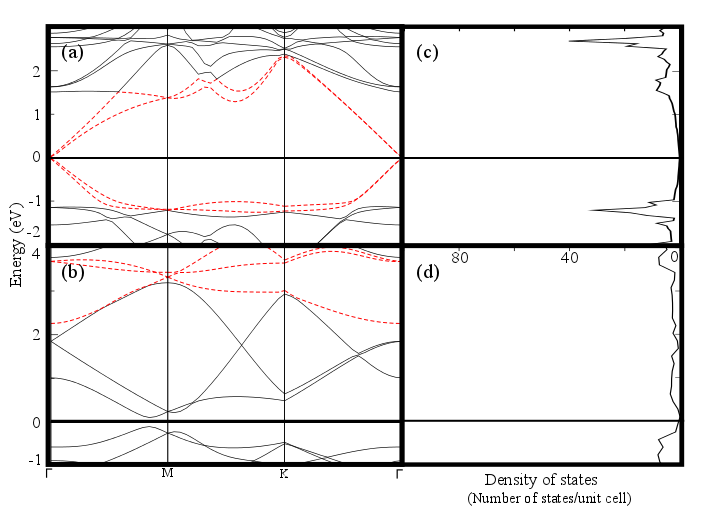}}
\caption{(Color online) Energy band structures of monolayer graphene on (a) Si-terminated quartz, (b) O-terminated quartz and their respective total densities of states (c) and (d). The Fermi energy is set to zero. The bands with a significant contribution from carbon are marked in red.\label{fig2}} 
\end{figure}

We undertook a DFT study of graphene on $\text{SiO}_2$ (alpha quartz) and $\text{Al}_2\text{O}_3$ (alpha sapphire or alumina), each with two surface terminations. Detailed description of the  calculations is given elsewhere \cite{jadaun}. The starting bulk structures for quartz and alumina were taken from literature. The lattice constants for bulk were then optimized and the resulting bulk unit cells were used for further calculations. Super cells of graphene on oxides were built using repeated layers of the bulk unit cell terminated by hydrogen on one end and capped by a graphene layer on the other, refer to Fig. \ref{fig1}. We found that $6*d_{\text{C−C}}$ graphene, (where $d_{\text{C−C}} = 1.42\AA$), was nearly commensurate with the hexagonal surface of the oxides, as shown in Fig. \ref{fig1}. Lattice mismatch between quartz and alumina and graphene was as low as ~0.19\% and ~0.42\%, respectively. The supercell structures were then allowed to relax.

\begin{figure}[ht!]
\scalebox{0.3}{\includegraphics{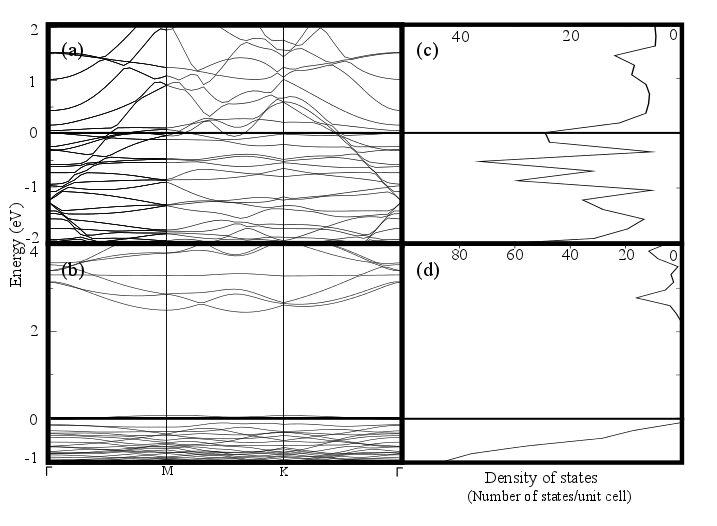}}
\caption{(Color online) Energy band structures of monolayer graphene on (a) Al-terminated alumina, (b) O-terminated alumina and their respective total densities of states in (c) and (d). The Fermi energy is set to zero.\label{fig3}} 
\end{figure}

For Si-terminated quartz, it was seen that graphene monolayer is pushed to an equilibrium distance of $3.0\AA$ showing very little in-plane rearrangement. The top few layers of Si and O surface were pushed away from graphene. There was no bonding between the Si surface and the C atoms, as a result of which, graphene retained its linear band structure, as shown in Fig. \ref{fig2}. We also noted that there was no charge transfer between the oxide and graphene sheet, refer to Fig. \ref{fig5}. Oxygen-terminated quartz however displayed very different properties. The equilibrium distance of graphene from the top surface was a much smaller $1.76\AA$, with graphene showing significant structural distortion and charge transfer to the oxide, as displayed in Fig. \ref{fig5}. It could be clearly seen that carbon atoms bonded with oxygen atoms and there were distinct signs of hybridisation between C-p and O-p orbitals. Dirac cone in graphene was destroyed, refer to Fig. \ref{fig2}. Parallel work done by Shemella et. al. \cite{shemella} also quoted similar results.

\begin{figure}[ht!]
\scalebox{0.3}{\includegraphics{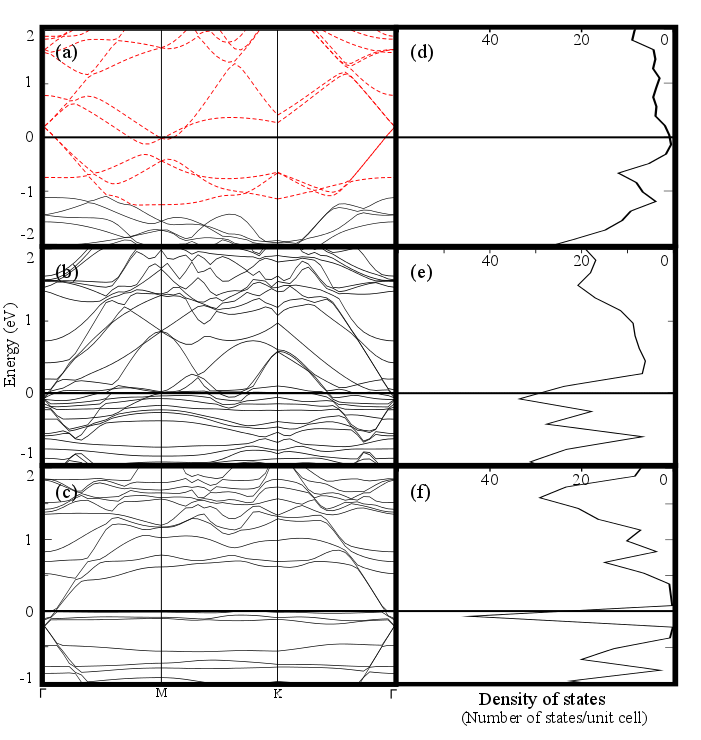}}
\caption{(Color online) Energy band structures of bilayer graphene on (a) O-terminated quartz, (b) Al-terminated alumina, (c) O-terminated alumina and their respective total densities of states in (d)–(f). The bands with a significant contribution from carbon are marked in red and the Fermi energy is set to zero.\label{fig4}} 
\end{figure}

Graphene on top of aluminium-terminated alumina showed considerable in-plane distortion relaxing to an equilibrium separation of $2.7\AA$. Carbon atoms bonded to the aluminium atoms of the surface and lost their linear band spectrum, refer to Fig. \ref{fig3}. For O-terminated alumina the result was distortion of graphene's spectrum too, however this time carbon bonded to the terminating O atoms of the alumina surface, as can be seen in Fig. \ref{fig3}. Graphene adopted an equilibrium distance of $2.15\AA$ losing some charge that moved to the oxide, shown in Fig. \ref{fig5}. Like the case of O-terminated quartz, there was significant hybridisation seen between the C-p and O-p orbitals signifying bonding. In analogy with the study of graphene on SiC, we added another layer of graphene on top, to see if that helped us recover the linear spectrum of graphene. If the lower graphene layer could act as a buffer and shield the layer above from interactions with the oxide below, one could hope to see the Dirac cone emerge. This behavior was seen for O-terminated quartz only, as seen in Fig. \ref{fig4}. For both terminations of alumina, linearity was not restored and additional layers of graphene would probably be required for achieving the same result.

\section{Edge effects}
\label{edge}
An accessible way of opening a band gap in graphene is to confine it to form graphene nanoribbons (GNRs).  This introduces edges to the graphene layer. The structure and nature of the edges alter the properties of graphene. Tight binding studies predict that GNRs can be either metallic or semiconducting, depending on the crystallographic direction of the nanoribbon axis. It was shown that GNRs with zigzag edges are always metallic\cite{nakada} whereas armchair nanoribbons can be either semiconducting or metallic as their band gap is an oscillating function of their width\cite{ezawa}.
\\Barone et. al. \cite{barone} conducted a DFT study of graphene nanoribbons of varying widths, both bare and hydrogen-terminated. For armchair nanoribbons, the band gap oscillated as a function of the ribbon's width. These oscillations were amplified upon termination of the edges by hydrogen, specially for narrow ribbons. While the tight binding calculations predicted the armchair GNRs (AGNRs) to be metallic for some certain widths, the DFT calculations predicted them to always be semiconduting. There was a periodicity of 3 seen in the band gap oscillations with width. They found that armchair GNRs with widths of 2-3 nm produced band gaps similar to those seen in Ge, while Si-like band gaps were seen in narrower GNRs of width 1-2 nm. They also studied H-terminated chiral nanoribbons, i.e., nanoribbons with low symmetry edges, as against armchair or zigzag that have high symmetry edges. It was noted that the amplitude of band gap oscillations decreased with increase in the chiral angle. For a chiral angle as low as $9^o$, the band gap oscillations became negligible. They thus suggested, that in order to engineer the band gap of GNRs, one needs to control the width as well as the chiral angle. Due to the strong dependence of band gap on the structure of GNRs, they also suggested using optical spectra of GNRs to characterise them. 
\\Similarly, a self-consistent pseudopotential method based study by Son et. al. \cite{son} showed that for hydrogen-passivated GNRs, both armchair and zigzag nanoribbons have non zero and direct band gaps. They found armchair GNRs to be semiconducting with the band gap decreasing as the width increases. As mentioned by Barone et. al. \cite{barone}, they too did not find any metallic AGNRs. This deviation from tight binding calculations, which had predicted some metallic AGNRs, was pinned down to be a result of the edge passivation of GNRs by hydrogen. The bonding of edge carbons to hydrogen effectively altered the on-site energy of those edge carbon atoms, a fact not taken into account in tight binding calculations. This passivation also changed the C-C bond lengths close to the edges. The alteration of C-C bond lengths, in turn, affected the amount of gap opening in a GNR. Zigzag graphene nanoribbons (ZGNRs) were also found to have direct band gaps which decreased with increasing width. The edge states in a ZGNR, however, were unique in that they displayed a very high density of states around the Fermi level, contrary to that seen in infinite graphene. This, in turn, led the ZGNRs to have magnetically polarised edge states, a property of great promise for spin transport. The ZGNRs possessed a magnetic insulating ground state with the spins aligned in a ferromagnetic pattern at a zigzag edge and anti-parallel ordering between opposite edges. The small spin-orbit coupling of carbon atoms was neglected in these calculations. The energy gaps in ZGNRs were said to arise from the staggered sublattice potentials due to the arrangement of spins on the edges of a ZGNR. As the width of a ZGNR increased, the strength of the staggered potential in the bulk of the GNR decreased, thereby decreasing the band gap.

\begin{figure}[ht!]
\scalebox{0.2}{\includegraphics{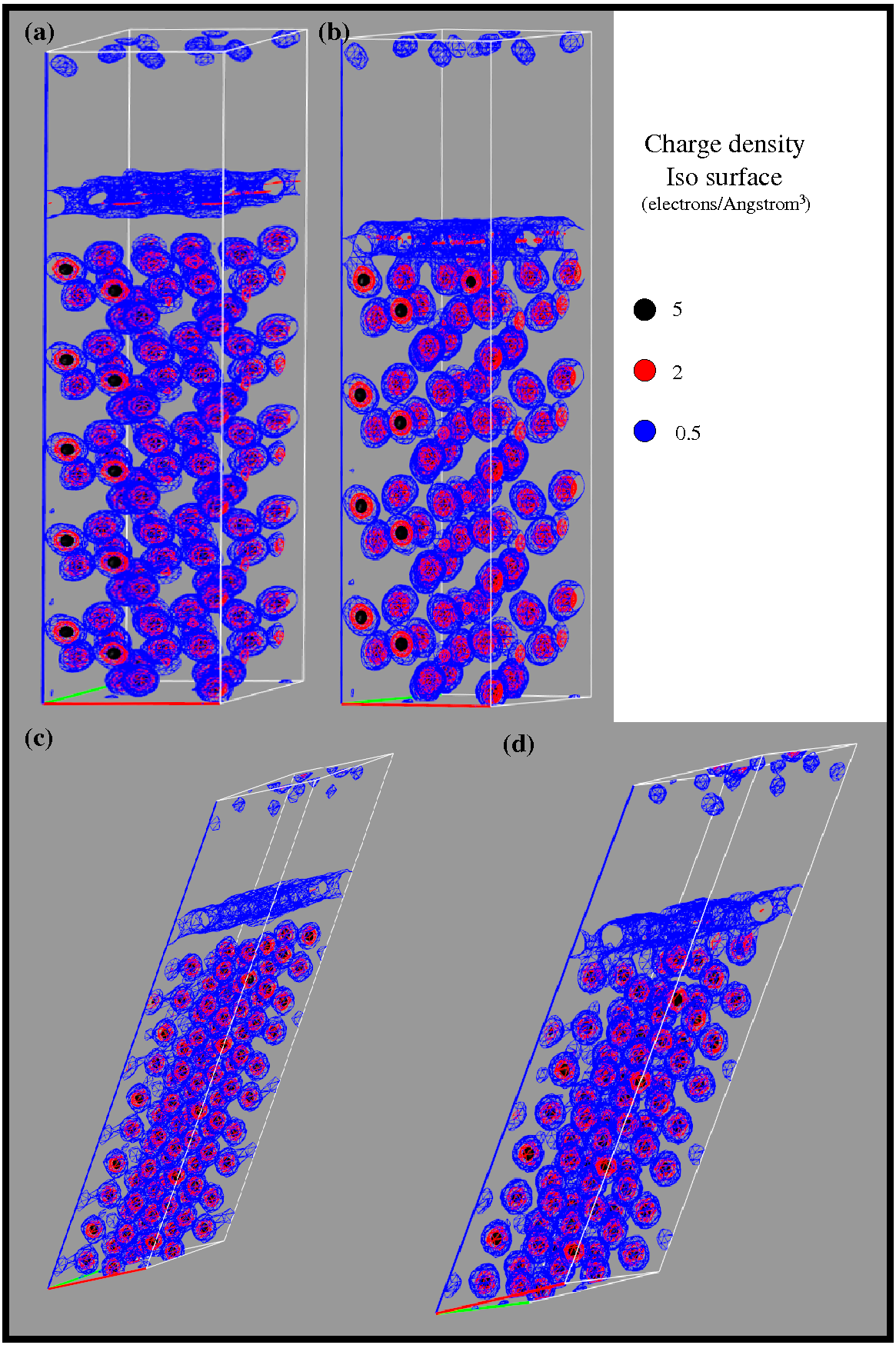}}
\caption{(Color online) Charge density plots for monolayer graphene on (a) Si-terminated quartz, (b) O-terminated quartz, (c) Al-terminated alumina and (d) O-terminated alumina.\label{fig5}} 
\end{figure}

In reality though, GNRs used for technology would have to be finite in length, even as they are confined along the width. Hod et. al. \cite{hod} conducted a study to gauge finite-size effects in zero-dimensional armchair GNRs. They showed that for GNRs with 12nm length, the density of state features were very different from what one would expect out of infinite GNRs. At a length of 38nm, one could clearly see characteristic features like Van-Hove singularities, constant density of states near the Fermi energy and a dip in the density of states at the Fermi energy. This behavior corroborated better with infinite length GNRs. On the whole, 40nm was a short enough length to see finite-size properties approaching infinite length GNR results. There were some edge effects seen close to the Fermi level even for longer GNRs but they died down as the length increased, becoming negligible only at micrometer lengths. This emphasized the importance of studying edge effects of length confinement, along with width, for realistic device simulations. Shemella et. al. \cite{shemella1} also undertook a similar study and made an important distinction between 1D (infinite length) AGNRs and 0D (finite length) AGNRs. 0D AGNRs possess zigzag edges along the width whereas 1D AGNRs do not. Much like 1D ZGNRs, zigzag edges for 0D AGNRs displayed magnetic behavior in the ground state with a very high density of state near the Fermi level. These localised spin moments at the edges provided a stagerred sublattice potential, which could open up a band gap for the nanoribbon. They found however that for non-metallic AGNRs the highest occupied molecular orbital (HOMO) and the lowest unoccupied molecular orbital (LUMO) were localised at the edges, whereas for metallic AGNRs they are delocalised throughout the nanoribbon. This caused the bandgap in metallic AGNRs to depend on the length of the nanoribbon while the bandgap in semiconducting AGNRs remained independent. Zhang et. al. \cite{zhang} performed a DFT calculation of hydrogen-passivated zigzag graphene nanoribbons (H-ZGNRs) atop O-terminated $\text{SiO}_2$ substrate. They found both the surfaces of ZGNRs as well as $\text{SiO}_2$ were distorted with the equilibrium distance between them being 1.38$\AA$. In the more stable O1 surface configuration, C-O bonds were formed but these helped pin the edges of H-ZGNRs, thereby suppressing the effect of H passivation. This, in turn, allowed the H-ZGNRs to be metallic in line with the tight binding predictions quoted earier. H-ZGNRs supported on the O2 surface also displayed C-O bonding with band gaps emerging in some cases, depending on the edge states and nanoribbon widths.

\section{Conclusion}
\label{conc}
To summarise, we have presented a review of theoretical studies, based on density functional theory, undertaken to understand graphene's interaction with variables in its surroundings. These include substrates, dielectrics and graphene edges. We particularly focussed on our work involving the impact of crystalline dielectric oxides on graphene's properties, specifically that on quartz and alumina. We described the role played by SiC, O-terminated quartz and alumina in perturbing graphene and opening a band gap in its spectrum. Introducing edges in graphene, by the process of confining it to make infinite or finite nanoribbons also, in many cases, caused a non-zero band gap to emerge. At a length of 40nm finite GNRs approached the behavior predicted for infinite GNRs retaining some edge effects that became negligible at the length of microns. 

\section{Acknowledgement}
\label{ack}
This work was supported by the NRI SWAN program. The authors acknowledge the allocation of computing time on NSF Teragrid machines Ranger (TG-DMR080016N) and Lonestar at the Texas Advanced Computing Center.


\begin{thebibliography}{18}
\bibitem{novoselov} K.S. Novoselov, A.K. Geim, S.V. Morozov, D. Jiang, M.I. Katsnelson, I.V. Grigorieva, S.V. Dubonos, A.A. Firsov, Nature 438 (2005) 197-200.
\bibitem{geim} A.K. Geim, K.S. Novoselov, Nature Materials 6 (2007) 183 - 191. 
\bibitem{zhou} S.Y. Zhou, G.-H. Gweon, A.V. Fedorov, P.N. First, W.A. de Heer, D.-H. Lee, F. Guinea, A.H. Castro Neto, A. Lanzara, Nature Materials 6 (2007) 770 - 775.
\bibitem{shemella} P. Shemella, S.K. Nayak, Appl. Phys. Lett. 94 (2009) 032101.
\bibitem{liao} L. Liao, X. Duan, Materials Science and Engineering: R: Reports 70 (2010) 354-370.
\bibitem{puls} C.P. Puls et. al., Appl. Phys. Lett. 99 (2011) 013103.
\bibitem{jadaun} P. Jadaun, S.K. Banerjee, L.F. Register, B. Sahu, J. Phys.: Condens. Matter 23 (2011) 505503.
\bibitem{varchon} F. Varchon, R. Feng, J. Hass, X. Li, B. Ngoc Nguyen, C. Naud, P. Mallet, J.-Y. Veuillen, C. Berger, E.H. Conrad, L. Magaud, Phys. Rev. Lett. 99 (2007) 126805.
\bibitem{matta} A. Mattausch, O. Pankratov, Phys. Rev. Lett. 99 (2007) 076802.
\bibitem{xu} Y. Xu, K.T. He, S.W. Schmucker, Z. Guo, J.C. Koepke, J.D. Wood, J.W. Lyding, N.R. Aluru, Nano Letters 11 (2011) 2735-2742.
\bibitem{giovannetti} G. Giovannetti, P.A. Khomyakov, G. Brocks, P.J. Kelly, J. van den Brink, Phys. Rev. B 76 (2007) 073103.
\bibitem{nakada} K. Nakada, M. Fujita, G. Dresselhaus, M.S. Dresselhaus, Phys. Rev. B 54 (1996) 17954–17961.
\bibitem{ezawa} M. Ezawa, Phys. Rev. B 73 (2006) 045432.
\bibitem{barone} V. Barone, O. Hod, G.E. Scuseria, Nano Lett. 6 (2006) 2748–2754.
\bibitem{son} Y.-W. Son, M.L. Cohen, S.G. Louie, Phys. Rev. Lett. 97 (2006) 216803.
\bibitem{hod} O. Hod, J.E. Peralta, G.E. Scuseria, Phys. Rev. B 76 (2007) 233401.
\bibitem{shemella1} P. Shemella, Y. Zhang, M. Mailman, P.M. Ajayan, S.K. Nayak, App. Phys. Lett. 91 (2007) 042101.
\bibitem{zhang} D.M. Zhang, Z. Li, J.F. Zhong, L. Miao, J.J. Jiang, Nanotechnology 22 (2011) 265702.
\end{thebibliography}
\end{document}